\documentclass[journal]{IEEEtran}

\usepackage{lineno} 

\usepackage{amsmath,amsfonts}
\usepackage{array}
\usepackage[caption=false,font=normalsize,labelfont=sf,textfont=sf]{subfig}
\usepackage{textcomp}
\usepackage{stfloats}
\usepackage{url}
\usepackage{verbatim}
\usepackage{graphicx}
\usepackage{cite}

\usepackage[bookmarksnumbered=true]{hyperref}
\hypersetup{
hidelinks,
colorlinks=true,
linkcolor=blue,
citecolor=blue,
urlcolor=black
}
\usepackage[linesnumbered,ruled,vlined]{algorithm2e}
\usepackage{bbm} 

\usepackage{amsthm} 

\usepackage{hyperref}
\usepackage{xcolor}
\hypersetup{
    colorlinks=true,      
    urlcolor=red          
}

\usepackage{enumerate} 

\begin{document}

\newcounter{probcounter}
\newcommand{\refporbcounter}[1]{\refstepcounter{probcounter}\theprobcounter \label{#1}} 

\title{Generative AI Enabled Matching for 6G Multiple Access
}

\author{
Xudong Wang, 
Hongyang Du,
Dusit Niyato,~\IEEEmembership{Fellow,~IEEE,}
Lijie Zhou,
Lei Feng,~\IEEEmembership{Member,~IEEE,} \\
Zhixiang Yang,
Fanqin Zhou,~\IEEEmembership{Member,~IEEE,}
Wenjing Li,~\IEEEmembership{Member,~IEEE}

\thanks{\textit{Corresponding author: Lei Feng.}}
\thanks{
X. Wang, L. Zhou, L. Feng, F. Zhou, Z. Yang and W. Li are with the State Key Laboratory of Networking and Switching Technology, Beijing University of Posts and Telecommunications, Beijing, China, 100876 
(e-mail: xdwang@bupt.edu.cn, 2024110893@bupt.edu.cn, fenglei@bupt.edu.cn, yangzx@bupt.edu.cn, fqzhou2012@bupt.edu.cn, wjli@bupt.edu.cn).}
\thanks{Hongyang Du is with the Department of Electrical and Electronic Engineering, University of Hong Kong, Pok Fu Lam, Hong Kong SAR, China (e-mail: duhy@eee.hku.hk).}
\thanks{Dusit Niyato is with the College of Computing and Data Science,
Nanyang Technological University, Singapore (e-mail: dniyato@ntu.edu.sg).}
}



\maketitle

\begin{abstract}

In wireless networks, applying deep learning models to solve matching problems between different entities has become a mainstream and effective approach. However, the complex network topology in 6G multiple access presents significant challenges for the real-time performance and stability of matching generation. Generative artificial intelligence (GenAI) has demonstrated strong capabilities in graph feature extraction, exploration, and generation, offering potential for graph-structured matching generation. In this paper, we propose a GenAI-enabled matching generation framework to support 6G multiple access. Specifically, we first summarize the classical matching theory, discuss common GenAI models and applications from the perspective of matching generation. Then, we propose a framework based on generative diffusion models (GDMs) that iteratively denoises toward reward maximization to generate a matching strategy that meets specific requirements. Experimental results show that, compared to decision-based AI approaches, our framework can generate more effective matching strategies based on given conditions and predefined rewards, helping to solve complex problems in 6G multiple access, such as task allocation.

\end{abstract}


\section{Introduction}

The matching problem is an important branch of combinatorial optimization, typically studied in graph theory. It has extensive research and applications in real-world scenarios, such as task scheduling, market matching, network design, and resource allocation. For example, in the transportation and supply demand field, passengers are matched with suitable vehicles and routes to maximize potential expected revenue~\cite{zhou2021graph}.
In market matching, determining how to match buyers and sellers to achieve optimal transactions directly affects the efficiency and fairness of the market. 
The widespread application of the matching problem demonstrates its value as a crucial method for addressing practical challenges, enhancing resource utilization, and promoting sustainable social development.

In the context of 6G multiple access, such as non-orthogonal multiple access (NOMA) and rate-splitting multiple access (RSMA), the importance of studying the matching problem is self-evident due to the needs for user coordination and interference management within the network. For instance, in NOMA-assisted wireless networks, user equipment and channel resources can be abstracted as nodes, and their allocation relationships as edges. Efficient matching methods can thus be used to manage and reduce interference, improving overall communication quality and system capacity \cite{10417791}. Therefore, applying matching methods can serve as a crucial tool for resource management, performance optimization, and interference coordination.
A considerable body of literature has delved into the matching problem in wireless networks, primarily focusing on stable matching and those methods that generate adaptive matching strategies using machine learning and deep learning models. However, these traditional approaches often face the following challenges and limitations.
\begin{itemize}
    \item \textbf{High Complexity}. When dealing with large-scale and highly complex systems (such as RSMA networks or edge computing systems), the complexity of traditional matching algorithms increases sharply due to the large number of players involved in the matching process, making real-time solutions unattainable~\cite{7105641}.
    \item \textbf{Dependence on complete information}. Stable matching and deep learning-aided methods typically assume that participants can provide a complete and fixed preference list~\cite{8382166}. However, in practice, participants may struggle to fully express their preferences, making traditional methods that rely on complete information difficult to apply.
    \item \textbf{Low exploration and convergence speed}. Decision-making AI, such as deep reinforcement learning (DRL), typically rely on repeated exploration and feedback for policy optimization, resulting in a training process that often requires numerous iterations to find a stable matching strategy~\cite{du2024diffusion}.
\end{itemize}


Generative artificial intelligence (GenAI) can learn the distribution characteristics of a given graph and generate new graphs based on changes in external conditions. These graphs can be viewed as forms of matching, where nodes represent matching objects and edges represent associations. Therefore, utilizing GenAI to construct graph-based matching provides a new approach to addressing these challenges.
In 6G multiple access wireless networks, applying GenAI to generate matching strategies involves several key steps~\cite{9920219}. First, it is necessary to collect data on network status, user demands, and channel conditions, and preprocess this data to ensure its quality and the effectiveness of model training. Next, using this data, train the selected generative models, such as variational auto-encoders (VAEs), generative adversarial networks (GANs), Transformers, and generative diffusion models (GDMs), enabling them to learn the latent associations between nodes and generate matching strategies based on graph theory. Using the trained generative models, generate new matching graphs according to the current network topology and user characteristics. 


The matching results represented by these graphs plays a crucial role in the design and optimization of multiple access networks.
On one hand, it can flexibly and efficiently manage interference and decode user noise to minimize inter-user interference, thereby enhancing system reliability and communication rate performance. On the other hand, match generation can optimize resource allocation and adjust network load to enhance communication network performance and achieve load balancing. These benefits are essential for creating more efficient and reliable wireless networks.

Considering the widespread application of matching problems in multiple access and the potential solutions offered by GenAI, this paper explores the application of GenAI-assisted matching generation methods in 6G multiple access wireless networks. 
First, we extensively summarize the applications of matching problems in various fields such as task allocation and next generation multiple access.
Then, we analyze and discuss the main GenAI models used for matching generation. Finally, we propose a matching generation framework based on diffusion models and illustrate how this framework supports image generation and wireless transmission in RSMA networks through a case study. The main contributions of this paper are summarized as follow.
\begin{itemize}
    \item We discuss the basic matching theory, and introduce commonly used GenAI models centered around the goal of matching generation, including their implementation principles, strengths, and weaknesses. This foundational knowledge reviews how different GenAI models can be used to generate efficient matching strategies.
    \item We investigate the applications of matching generation in multiple areas such as potential drug targets prediction and task offloading in vehicular networks, providing a comprehensive summary and analyzing the advantages of GenAI-based matching methods in 6G multiple access schemes.
    \item We propose a effective matching generation framework based on generative diffusion models and demonstrate its effectiveness through a case study of AIGC service provider selection problem in RSMA-enabled wireless networks.
\end{itemize}

\section{Overview of Matching Theory and Generative AI Models}



\subsection{Overview of Basic Matching Theory}

Matching plays a crucial role in the efficient allocation of resources across various fields such as transportation planning, market operations, and wireless networks. Matching is generally defined as a process where multiple players are paired according to certain rules to achieve an optimal and stable state. The preference list is a core element of matching, reflecting the priorities or inclinations of users when seeking to pair with other players. For instance, in multiple access wireless networks, tasks and resources can be viewed as two types of players. Tasks may rank resources based on factors such as computing power, latency, and bandwidth, while resources may prioritize tasks according to their urgency, required resources, and priority. These rankings form respective preference lists, guiding the matching process toward an outcome that maximizes satisfaction or utility for all participants.
Based on the values of player quotas, matching problems are typically classified into the following categories:
\begin{itemize}
    \item One-to-one matching: Each player can only be matched with one player from another group. A classic application is the stable marriage problem. Another example is the service providers selection in metaverse services \cite{du2024diffusion}.
    \item Many-to-one matching: One group of players can be matched with multiple players from another group, but the latter can only choose one player. An example is the user access problem in NOMA networks, where one channel can serve multiple users, but each user can only choose one channel \cite{9352956}.
    \item Many-to-many matching: Each player can be matched with multiple players from another group simultaneously. For example, in vehicular networks, client vehicles offload tasks to multiple service vehicles \cite{liu2024dnn}.
\end{itemize}

To achieve stable matching results effectively, matching algorithms have been widely studied. The Deferred Acceptance (DA) algorithm is a significant stable matching algorithm, where one group of players iteratively pairs with another group based on their preference lists, and the other group rejects all players except for their most preferred offers. 
Other stable matching algorithms include Top Trading Cycles, the Boston Mechanism, and Maximal Weighted Matching. Apart from stable matching algorithms, with the recent development of machine learning and artificial intelligence, decision-making AI algorithms, such as DRL, have been applied to adaptively achieve matching by learning and optimizing strategies. These methods are particularly effective in scenarios requiring continuous decision-making and feedback \cite{9352956, 9964376}. However, these approaches often require complete preference lists, but many users are reluctant to provide detailed preference information, especially when sensitive data is involved. Moreover, as the number of users and available options (resources, service providers and etc.) increases, the scale of the matching problem grows exponentially, generating and collecting complete preference lists becomes impractical in large-scale networks. Against this backdrop, some matching methods based on GenAI have been explored. Next, we will briefly review the GenAI models for matching generation.

\vspace{-0.2cm}
\subsection{Overview of GenAI Models for Matching Generation}

\begin{figure*}
    \centering
    \includegraphics[width=0.92\linewidth]{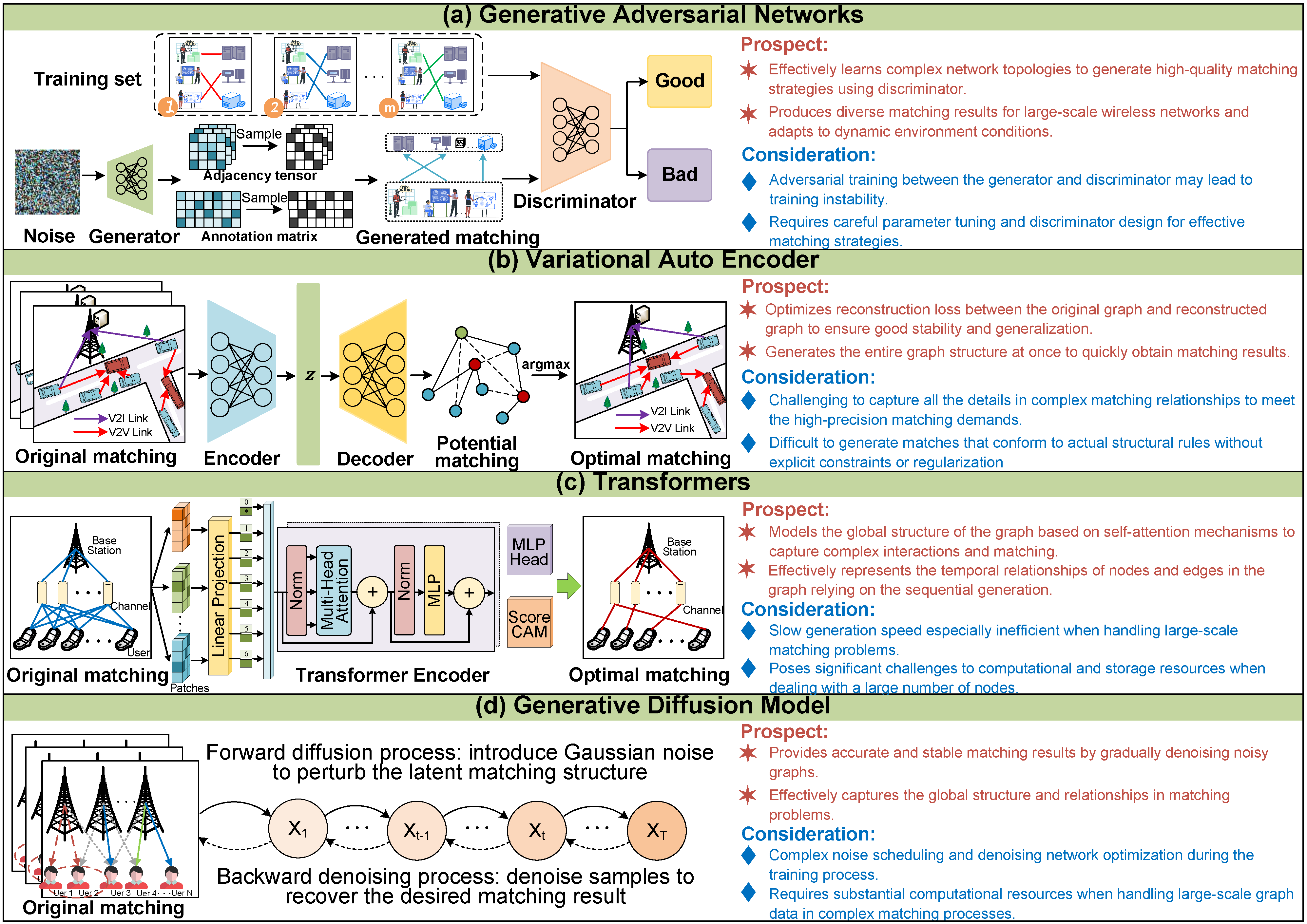}
    \caption{The summary of matching generation using GenAI models: Generative Adversarial Networks, Variational Auto Encoder, Transformer and Generative Diffusion Model. Conduct an in-depth investigation of four different GenAI models from the perspectives of principles, advantages, and disadvantages.}
    \label{GAI_model}
\end{figure*}

We discuss GenAI models that can be used for matching generation as follows (Fig.~\ref{GAI_model}).

\setlength{\abovecaptionskip}{-0.15cm}   

\subsubsection{\textbf{Generative Adversarial Networks}}GANs are a type of deep learning model consisting of two neural networks: a generator and a discriminator. The generator creates realistic graph-structured matching samples from random noise to deceive the discriminator, while the discriminator attempts to distinguish between real and generated samples. Through continuous competition and optimization, the quality of the generated samples improves to the point where the discriminator can no longer tell them apart. For instance, MolGAN~\cite{de2018molgan} implicitly generates desired molecular graphs without probabilities. The generator samples features of nodes and edges, and the discriminator differentiates between samples from the dataset and those from the generator. A reward network based on reinforcement learning drives the generator to produce molecules with various chemical properties.
For multiple access, GANs can generate diverse matching strategies through adversarial competition, adapting to complex and dynamic network conditions and multi-user access requirements. 
However, the dynamic competition between the generator and discriminator is essentially a zero-sum game, which can lead to unstable model training. Careful balancing of training and adjusting loss functions is necessary to achieve stability.

\subsubsection{\textbf{Variational Auto-encoder}}VAEs comprises an encoder and a decoder, where the encoder encodes input data into a probabilistic representation of the latent space, while the decoder generates reconstructed data from this distribution. By minimizing the reconstruction loss and the Kullback-Leibler (KL) divergence loss, the VAEs decoder can generate a fully connected graph with probabilistic representations to achieve an optimal graph structure \cite{simonovsky2018graphvae}.
Inspired by this approach, the adjacency matrix representing node connections in the encoder can be designed as an antisymmetric block matrix, with elements on the main diagonal set to zero. 
Additionally, an antisymmetric block matrix generation constraint can be introduced in the decoder to ensure that the generated structure adheres to a bipartite graph suitable for optimal matching strategies. 
By optimizing the reconstruction loss and KL divergence, the VAEs-based matching method can exhibit good stability and generalization.
This allows for smooth interpolation within the continuous latent space, ensuring diversity in the generated graphs and the exploration of various matching possibilities. The ability to generate the entire graph structure in one go improves the speed and efficiency of matching generation, making it particularly suitable for 6G multiple access networks that require rapid acquisition of optimal pairings. However, due to the limitations of the loss function, the reconstruction quality of VAEs might not meet the demands of large-scale, high-precision matching applications.

\subsubsection{\textbf{Transformers}}Transformers utilize the encoder composed of self-attention mechanisms and feedforward networks to learn and capture global dependencies between matching entities through linear projection. Then they decode these dependencies through cross-attention layers to generate matching relationships. For instance, Graph Transformer Networks in \cite{yun2019graph} use graph transformation layers to dynamically generate meta-paths composed of multiple nodes and edges to capture useful connections. Specifically, it constructs an adjacency matrix and applies graph convolution operations to extract the feature information of nodes, thereby forming the final graph structure that includes nodes and connection relationships. With the global dependency modeling, Transformers are effectively suited for complex multidimensional matching problems. Additionally, the sequential decoding mechanism of Transformers helps capture the temporal relationships between nodes and edges in matching. However, sequential generation can result in slower processes, especially when handling large matching problems. Moreover, the high computing complexity associated with the self-attention mechanism can pose significant challenges in handling matching problems in large-scale multiple access networks.

\subsubsection{\textbf{Generative Diffusion Model}}GDMs draw on principles of nonequilibrium thermodynamics to generate diverse graph structures that represent matching strategies through a process of gradually adding noise to the original graph samples and then denoising them in reverse. GDMs are probabilistic models that consist of two stages: diffusion and denoising. Initially, the model introduces Gaussian noise step-by-step to perturb the latent matching structure. Then, it iteratively denoises samples from a normal distribution to obtain the desired feature distribution. GDMs have recently been widely applied to high-quality data generation tasks such as image generation and audio synthesis. For example, DiGress~\cite{vignac2022digress} uses a discrete diffusion process to edit graph noise and generate graphs with classified node and edge attributes, showing excellent performance in handling drug sample datasets. In the context of 6G multiple access, GDMs can provide accurate and stable matching schemes between devices and dynamically changing channels, thereby improving communication efficiency and reducing network fluctuations. However, due to the complexity of noise scheduling and denoising network optimization involved in the training process, model iteration requires substantial computational resources.

Fig.~\ref{GAI_model} summarizes the methods mentioned above for matching generation using various GenAI models. Whether through one-step generation or iterative inference, each model has its unique features and limitations. The use of these generative AI models in generating matching strategies allows for the effective handling of dynamic and heterogeneous network conditions in 6G multiple access wireless networks, leading to optimized resource allocation, interference management, and overall network performance improvement.

\section{Applications of GenAI Enabled Matching}

\subsection{Applications of Generative AI Enabled Matching}
We discuss the applications of GenAI-enabled matching in the areas of potential drug target prediction, metaverse services and vehicular networks, which are shown in Fig.~\ref{applocation} (a-c).

\subsubsection{Potential Drug Targets Matching}The global pandemic caused by Coronavirus Disease 2019 (COVID-19) has highlighted the critical importance of repurposing drugs with proven safety and no toxic side effects for treating patients. To explore drug-coronavirus-host protein-protein interaction (PPI) matching, the authors in \cite{ray2020predicting} input the adjacency matrix of nodes and the feature matrix of low-dimensional embeddings into a VAE model for training. The trained model was then applied to a drug-virus-host node set without edges to identify the most likely matches. By identifying the most probable connections among 992 drugs and 78 CoV host proteins, the approach demonstrated its effectiveness in discovering targeted drug therapies. 

\subsubsection{Metaverse Services}To ensure that users in a human-centric Metaverse can access AI-generated content (AIGC) services and obtain quality of experience (QoE) for users, the authors in~\cite{du2024diffusion} proposed an AI-generated optimal decision (AGOD) algorithm to achieve optimal matching between service provider (SP) and users. To address this combinatorial optimization problem with discrete variables, GDMs gradually add noise to the current optimal allocation in the environment. During the inverse reasoning phase, the optimal decision generation network acts as a denoiser, iteratively restoring the optimal matching strategy based on environmental conditions.

\begin{figure*}
    \centering
    \includegraphics[width=0.9\linewidth]{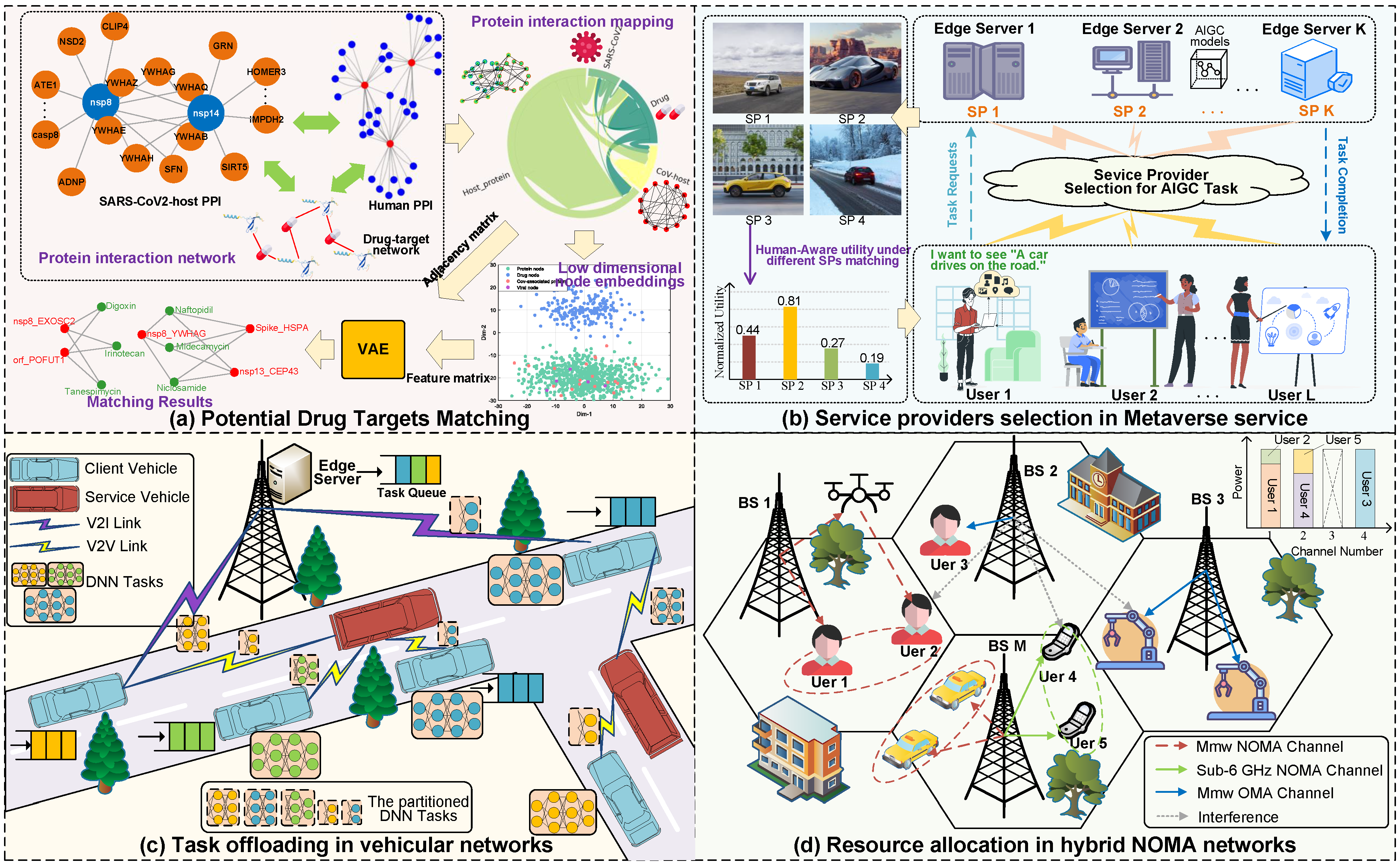}
    \caption{The application of matching in potential drug targets prediction, metaverse service, vehicular networks, and hybrid NOMA networks. In potential drug targets prediction, a VAE model is applied to identify the optimal matching between drugs, viruses, and hosts, effectively exploring the potential for targeted drug therapies~\cite{ray2020predicting}. In metaverse service, efficient matching between users and edge servers can enhance the immersive experience of users~\cite{du2024diffusion}. In vehicular networks, service vehicles, and base station are considered vertices, while V2V and V2I links for DNN task offloading are considered edges~\cite{liu2024dnn}. A structured matching strategy greatly enhances task completion efficiency. In multiple access wireless networks, decision-making AI algorithms such as DRL interact with the dynamic environment and maximize the reward function to achieve effective matching between physical entities and wireless resources, thereby improving system spectral efficiency and throughput~\cite{9352956, 9964376}.}
    \label{applocation}
\end{figure*}

\subsubsection{Vehicular Networks}To address the issue of task-resource matching in vehicular networks, the authors in \cite{liu2024dnn} introduced a multi-agent deep reinforcement learning algorithm based on GDMs, aimed at partitioning tasks integrated with deep neural networks (DNNs) and assigning them to appropriate vehicle-to-vehicle (V2V) and vehicle-to-infrstructure (V2I) systems to enhance task completion efficiency, as shown in Fig.~\ref{applocation}. Compared to DRL, diffusion models have demonstrated a stronger capability to interact and sense the dynamic environment of vehicular networks, thereby enhancing the action sampling efficiency during the matching process.

\vspace{-0.2cm}
\subsection{Lessons Learned}

The exploration of above matching applications has highlighted several critical insights and lessons.

\begin{itemize} 
    \item GenAI models, unlike traditional static optimization algorithms, can adapt optimization strategies in dynamic environments, making them ideal for real-time applications requiring fast, high-quality decisions. 
    \item Matching problems can be naturally represented as graphs, with nodes as elements (e.g., tasks, resources) and edges as relationships, allowing for intuitive visualization and capturing complex dependencies. 
    \item GenAI models excel at generating graph structures and pairing strategies that address real-world constraints, optimizing system performance in a scalable and real-time manner. 
\end{itemize}

These insights underscore strong adaptability of GenAI models in match generation, which has been applied in fields like potential drug targets prediction. In multiple access wireless networks, it is essential to explore optimal matching between physical entities, such as base stations (BSs) and user devices, and wireless resources like channels and bandwidth, as shown in Fig.~\ref{applocation} (d). The existing work primarily relies on traditional decision-based machine learning approaches such as DRL. For example, the authors in \cite{9352956} introduced an recurrent neural networks (RNN)-supported DRL framework for channel allocation in hybrid NOMA systems, aiming to enhance spectral efficiency and environmental adaptability. In \cite{9964376}, the authors integrated multi-agent deep reinforcement learning (MADRL) into a multi-cell hybrid NOMA scenario, where each BS is treated as an agent learning the three-dimensional association features among users, channels, and BS independently and without cooperation, leading to overall rate improvements. However, these traditional DRL methods face significant challenges in policy optimization and computing complexity under user overload~\cite{9625800}. Therefore, integrating GenAI into 6G multiple access for match generation to further enhance capabilities such as wireless connectivity is imperative.

\section{Matching Generation For 6G Multiple Access}

\subsection{Motivations and Challenges}

GenAI models show great adaptability in generating matching strategies for 6G multiple access networks. For example, GANs can quickly adapt to dynamic wireless networks through adversarial training, while GDMs excel in iterative denoising for precise matching in complex network conditions. Integrating GenAI with graph methods offers robust support for multiple access frameworks. However, applying GenAI for matching generation in 6G multiple access still faces various challenges.

\begin{itemize}
    \item \textit{Generation Quality and Computing Resources}: GenAI models typically require substantial computing resources and time for training, especially when iteratively generating solutions to complex matching problems in 6G multiple access. The high computing cost and training time may limit the efficiency and feasibility of GenAI models in practical applications.
    \item \textit{Adaptability to Dynamic Environments}: Wireless network environments are often dynamic, including changes in user demands, network conditions, and more. GenAI models need to be continuously updated and optimized to adapt to these changes; otherwise, they may generate unsuitable matching strategies.
    \item \textit{Scalability and Real-time Performance}: As the number of users and devices in multiple access wireless networks increases, the scale of the matching problem grows exponentially, requiring GDM models to handle larger graph structures and more complex relationships. This significantly increases the complexity of model training and inference, making it more difficult to generate high-quality matching strategies in real time.
\end{itemize}

\vspace{-0.2cm}
\subsection{The Proposed Framework}

Due to the potential advantages of GDMs in generation quality, scalability, and adaptability to dynamic environments, this paper proposes a conditional diffusion model-assisted deep generative framework to obtain optimal matching results for 6G multiple access networks, as shown in part A of Fig.~\ref{framework}. This framework consists of the following functions:

\begin{figure*}
    \centering
    \includegraphics[width=0.9\linewidth]{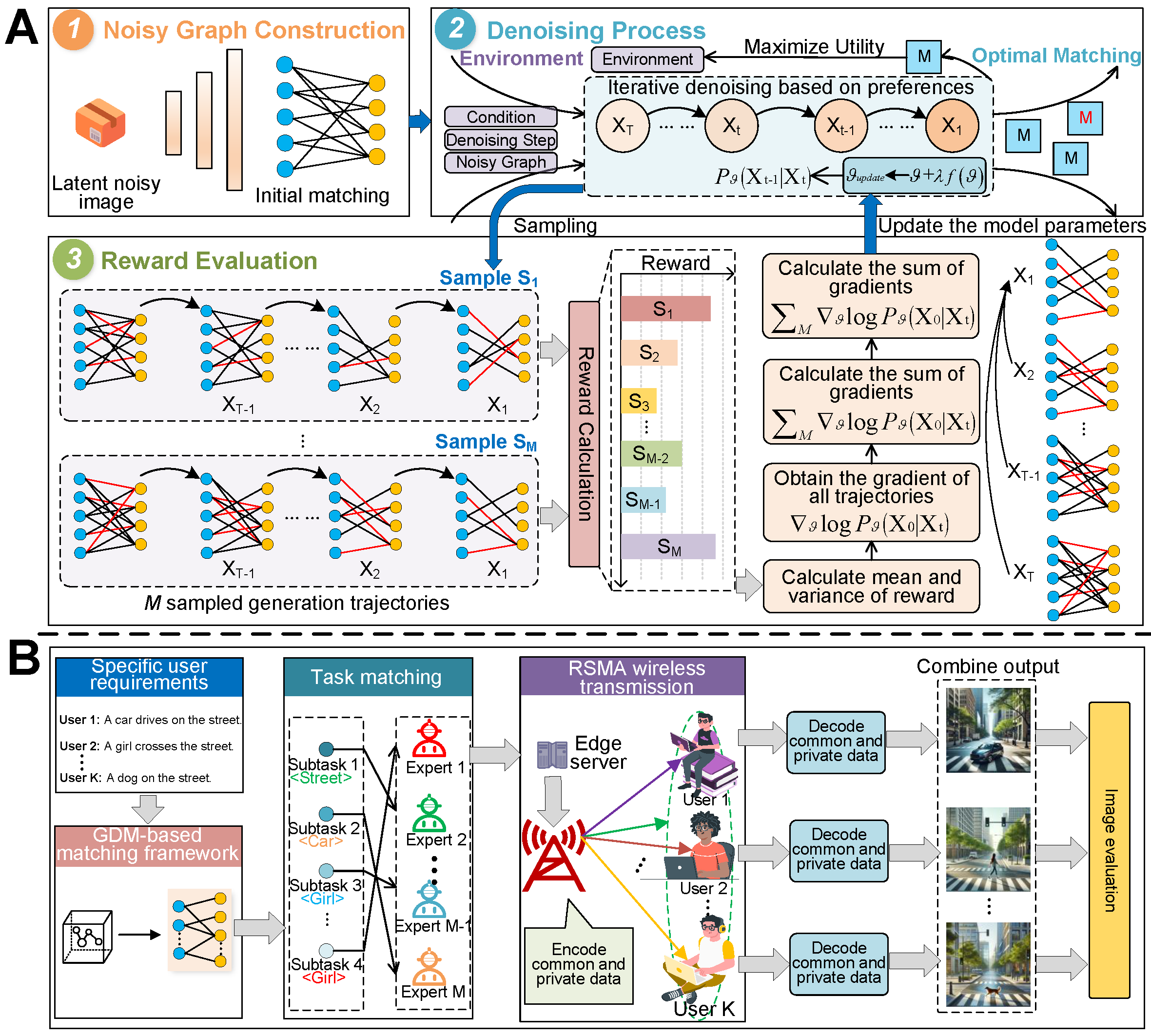}
    \caption{The framework of the proposed matching method. In part A, the framework follows four steps. Step 1, the random noise is mapped to a noisy graph through one-hot encoding. Step 2, the condition, denoising step, and the noisy graph containing matching relationships are input into the denoising network to iteratively generate the matching strategies. Step 3, multiple generated trajectories are sampled from the denoising network, and rewards are calculated to obtain the sum of gradients for model parameter updating. Finally, the denoising network which is iteratively updated can generate the optimal matching strategy based on the current environment. Part B illustrates the RSMA-aided AIGC service provider selection problem in Section~\ref{case_study}.}
    \label{framework}
\end{figure*}

\begin{itemize}
    \item \textbf{Step 1: Noisy Graph Construction}: Since raw noise cannot be directly applied to the discrete denoising process, one-hot encoding is first used to encode the noise into a noisy graph containing vertices and edges. This ensures that subsequent steps can naturally denoise the noise into the desired matching strategy represented by the graph structure. The noisy graph can be a fully connected graph or a random graph structure generated based on the characteristics of the problem.
    \item \textbf{Step 2: Denoising Process}: The environment condition (e.g., network status, user demands), the number of denoising steps, and the generated noisy graph are input into the denoising network. This network iteratively generates the required graph structure by adding or removing vertices or edges between vertices at each denoising step. In this context, vertices represent physical entities such as base stations and users, while edges represent the matching relationships between these entities. The denoising process is modeled as a Markov decision process (MDP), where the graph structure at each diffusion step represents the state, and the transformation of the graph structure corresponds to actions in the MDP. Therefore, the direction of denoising at each step is determined by the learned conditional probability.
    \item \textbf{Step 3: Reward Evaluation}: A reward function tailored to the specific matching problem is designed to assess the effectiveness of the generated matching strategy. The calculated reward value is input into the evaluation network to obtain the sum of gradients for the generation trajectory of the sampled  matching graph during the denoising process.
    \item \textbf{Step 4: Matching Strategy Generation}: Using the trained GDMs, graph structures are generated from the noisy graph through a reverse denoising process to determine the final matching pairs, thereby solving practical problems in 6G multiple access wireless networks.
\end{itemize}

The proposed diffusion model-based matching framework leverages iterative denoising to progressively generate matching solutions, effectively improving the matching accuracy compared to one-shot generation methods. Moreover, the Markov decision process during the denoising phase ensures that matching decisions can adapt in real-time to changing network conditions. Additionally, the use of graph-based structures allows for handling larger-scale problems while maintaining accuracy, addressing scalability challenges in extended 6G multiple access networks.

\section{Matching Generation For AIGC Service in RSMA-Aided Networks}
\label{case_study}

\subsection{Case Study}

\subsubsection{Experimental Configuration}Through an example, we demonstrate how to use the proposed matching generation framework to support AIGC services based on 6G RSMA.
As shown in part B of Fig.~\ref{framework}, we consider a setup with 15 users who have similar image generation needs and a server equipped with 6 AIGC models (called ``experts"). The server integrates diffusion models for image generation and our proposed diffusion model-based framework for optimal matching generation. The experts employ the first diffusion models for image generation to support AIGC services.
And the final image generated for each user is composed of outputs from several specific experts. Since users have different requirements, selecting different experts for AIGC services will result in varying QoE. The users request image generation from the server, with each expert on the server specializing in generating images of different styles, such as landscapes, cars, animals, and so on. The server then transmits the intermediate results generated by the experts to the users via the RSMA network, where the intermediate output results of shared experts as common data streams and the intermediate output results of user-specific experts as private data streams. Finally, users combine the intermediate results of the image content to ultimately produce the desired image. Since the selection and combination of different experts for image generation can impact user experience, we apply the proposed framework to generate matching strategies that guide the alignment between the expert components and users under various conditions, ensuring a customized service experience for the users. It is important to note that experts generating landscapes can also serve users that require animal images, although the quality of their experience may not be as optimal.

Considering that users are distributed uniformly within a range of $50\sim100~\mathrm{m}$, the signal frequency is set to $2.4~\mathrm{GHz}$, and wireless channel fading is taken into account. The signal-to-noise ratio (SNR) is set from $-10~\mathrm{dB}$ to $30~\mathrm{dB}$.
During the training, the generation conditions are set as the matching results between experts and users, aiming to maximize the total QoE of overall system while ensuring that computing resources and communication energy consumption remain within a certain range. 
Meanwhile, the intermediate images shared by users are encoded as common data streams, while the intermediate images unique to each user are encoded as private data streams. 
In addition, Zero-Forcing (ZF) beamforming is employed to mitigate interference between different data streams, thereby significantly improving image quality. In the proposed framework, our goal is to use a server equipped with experts to generate the desired parts of an image and wirelessly distribute them to users, who will then combine these parts to produce the final image. The final output of each user is an integration of images generated by multiple experts, and the quality of the generated image is assessed using a no-reference (or blind) image quality evaluation model, BRISQUE\footnote{\href{http://live.ece.utexas.edu/research/quality/BRISQUE_release.zip}{\textcolor{blue}{http://live.ece.utexas.edu/research/quality/BRISQUE\_release.zip}}}, which provides a clear evaluation score to measure QoE of users. 

\begin{figure}[!t]
    \centering
    \includegraphics[width=0.9\linewidth]{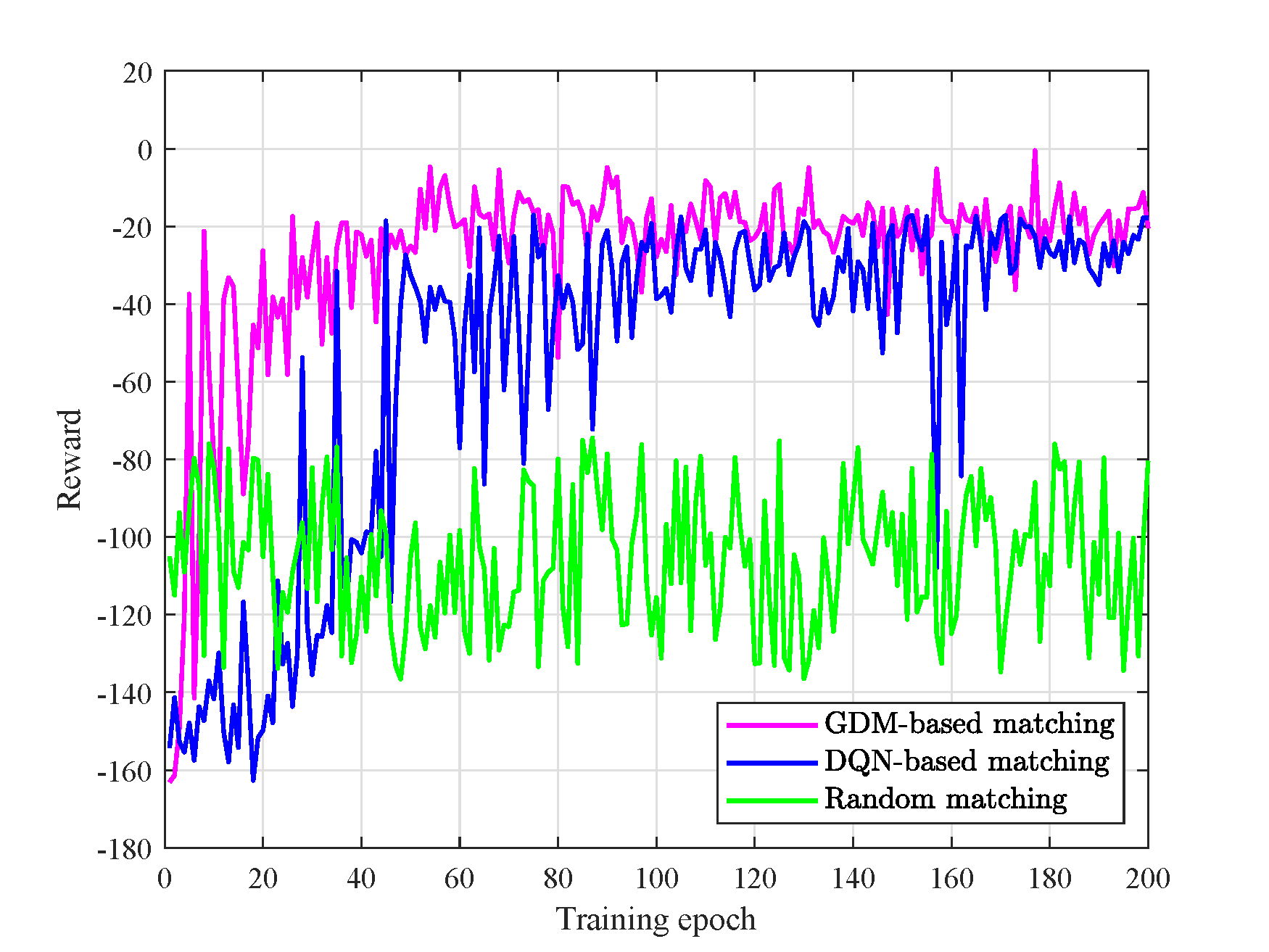}
    \caption{The training curve of the proposed GDM-based matching method and the comparison with other methods.}
    \label{convergence}
\end{figure}

\begin{figure}[!t]
    \centering
    \includegraphics[width=0.9\linewidth]{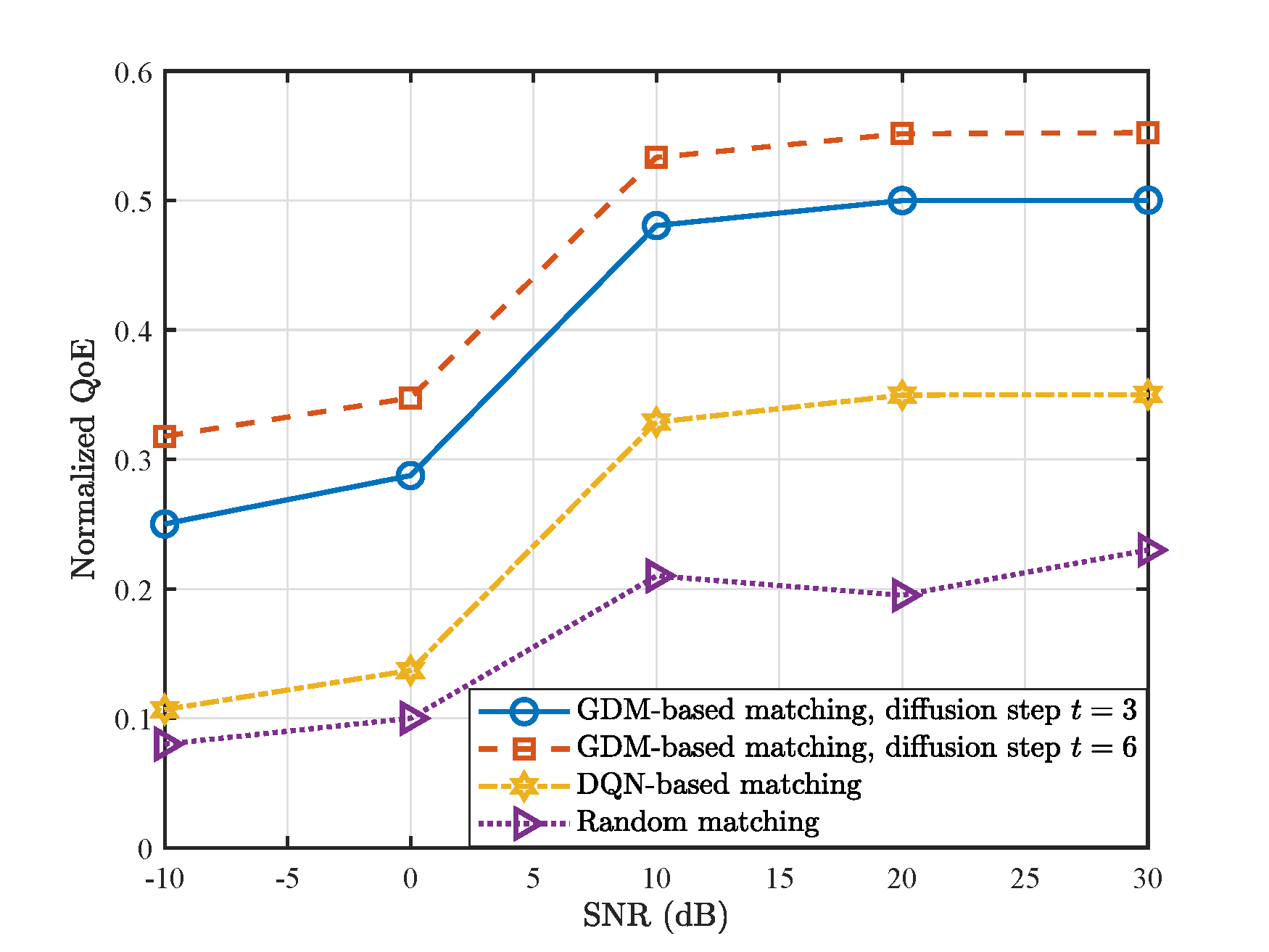}
    \caption{Normalized QoE versus SNR.}
    \label{Normalized_QoE_versus_SNR}
\end{figure}

\vspace{-0.4cm}
\subsection{Performance Analysis}

Fig.~\ref{convergence} shows the reward comparison between the proposed GDM-based matching method and other methods. First, it can be seen that the proposed method achieves an average reward of approximately -17 after 200 epochs, outperforming the deep Q-network (DQN) and random matching, which have average rewards of -35 and -100, respectively. This indicates that the diffusion model in the proposed matching generation framework can iteratively denoise towards reward maximization, generating optimized matching graphs. This allows the edge server at the base station to select the appropriate experts to perform image generation tasks, significantly improving the overall QoE performance. Moreover, the proposed method converges faster than the DQN method, demonstrating that our framework can effectively meet the large-scale access requirements of 6G multiple access networks.

Fig.~\ref{Normalized_QoE_versus_SNR} illustrates the normalized QoE versus SNR. First, it can be seen that the proposed method outperforms other methods at any SNR level, demonstrating its effectiveness in selecting the appropriate experts at the edge server to perform specific image generation tasks. Additionally, the performance of the proposed GDM-based matching method with 6 diffusion steps is better than that with 3 diffusion steps. This indicates that our matching generation framework can learn more latent features and produce more stable matching strategies by increasing the number of diffusion steps.

\vspace{-0.1cm}
\section{Future Directions}

\subsection{Multi-objective Matching Design}

In 6G networks, matching problems typically involve optimizing multiple objectives such as communication resource allocation, energy consumption, and latency. Matching generation methods that focus solely on a single objective are insufficient to meet the multidimensional demands of complex 6G scenarios. Future research could explore the design of diffusion models that adapt to multi-objective loss functions by incorporating trade-off mechanisms to balance various performance metrics, which will be key to fine-tuning resource management and matching optimization in 6G network.

\vspace{-0.5cm}
\subsection{Efficiency and Scalability}

As the scale of devices in 6G networks continue to grow, the complexity of matching problems increases exponentially. Future research could focus on developing lightweight and efficient diffusion model architectures to enhance computational efficiency in generating matching strategies. 
Additionally, employing techniques such as distributed computing or federated learning could help models quickly generate matching strategies in large-scale networks, avoiding computational bottlenecks. 
Research in this direction will help address high-load, multi-user scenarios frequently encountered in 6G networks, ensuring real-time matching generation and scalability.

\vspace{-0.5cm}
\subsection{Integrating Interactive Learning and Human-AI Collaboration}

Matching strategies in 6G networks involve complex user preferences, and by interacting with human experts, GenAI models can receive valuable insights to generate more practically effective matching strategies. Future research could focus on integrating human-AI collaboration into the learning process, adopting interactive AI models that dynamically adjust matching strategies while continually optimizing outcomes. This interactive learning approach will improve the applicability of matching generation, especially in highly uncertain or diverse-demand environments within 6G networks.

\section{CONCLUSION}

In this paper, we have investigated how GenAI models generate stable matching strategies for 6G multiple access. Specifically, we have first summarized the key points of matching theory and GenAI models, exploring the applications of matching generation in fields such as potential drug prediction and vehicular networks. Then, we have proposed a GDM-enabled matching method that guides the diffusion model to iteratively denoise toward reward maximization by evaluating a predefined reward function, thus solving the matching problem in 6G networks under given conditions. The effectiveness of the framework has been validated through a case study on selecting AIGC service providers in RSMA networks, providing valuable insights for addressing complex matching problems between entities in 6G multiple access.


\bibliographystyle{IEEEtran}
\bibliography{references}

\begin{thebibliography}{10}
\providecommand{\url}[1]{#1}
\csname url@samestyle\endcsname
\providecommand{\newblock}{\relax}
\providecommand{\bibinfo}[2]{#2}
\providecommand{\BIBentrySTDinterwordspacing}{\spaceskip=0pt\relax}
\providecommand{\BIBentryALTinterwordstretchfactor}{4}
\providecommand{\BIBentryALTinterwordspacing}{\spaceskip=\fontdimen2\font plus
\BIBentryALTinterwordstretchfactor\fontdimen3\font minus
  \fontdimen4\font\relax}
\providecommand{\BIBforeignlanguage}[2]{{%
\expandafter\ifx\csname l@#1\endcsname\relax
\typeout{** WARNING: IEEEtran.bst: No hyphenation pattern has been}%
\typeout{** loaded for the language `#1'. Using the pattern for}%
\typeout{** the default language instead.}%
\else
\language=\csname l@#1\endcsname
\fi
#2}}
\providecommand{\BIBdecl}{\relax}
\BIBdecl

\bibitem{zhou2021graph}
F.~Zhou, S.~Luo, X.~Qie, J.~Ye, and H.~Zhu, ``{Graph-based Equilibrium Metrics
  for Dynamic Supply--demand Systems with Applications to Ride-sourcing
  Platforms},'' \emph{Journal of the American Statistical Association}, vol.
  116, no. 536, pp. 1688--1699, 2021.

\bibitem{10417791}
M.~B. Singh, H.~Singh, and A.~Pratap, ``{Stable Matching Based Revenue
  Maximization for Federated Learning in UAV-Assisted WBANs},'' \emph{IEEE
  Transactions on Services Computing}, vol.~17, no.~4, pp. 1835--1846, 2024.

\bibitem{7105641}
Y.~Gu, W.~Saad, M.~Bennis, M.~Debbah, and Z.~Han, ``{Matching theory for future
  wireless networks: fundamentals and applications},'' \emph{IEEE
  Communications Magazine}, vol.~53, no.~5, pp. 52--59, 2015.

\bibitem{8382166}
Q.~Mao, F.~Hu, and Q.~Hao, ``{Deep Learning for Intelligent Wireless Networks:
  A Comprehensive Survey},'' \emph{IEEE Communications Surveys \& Tutorials},
  vol.~20, no.~4, pp. 2595--2621, 2018.

\bibitem{du2024diffusion}
H.~Du, Z.~Li, D.~Niyato, J.~Kang, Z.~Xiong, H.~Huang, and S.~Mao,
  ``{Diffusion-based Reinforcement Learning for Edge-enabled AI-generated
  Content Services},'' \emph{IEEE Transactions on Mobile Computing}, 2024.

\bibitem{9920219}
X.~Guo and L.~Zhao, ``{A Systematic Survey on Deep Generative Models for Graph
  Generation},'' \emph{IEEE Transactions on Pattern Analysis and Machine
  Intelligence}, vol.~45, no.~5, pp. 5370--5390, 2023.

\bibitem{9352956}
J.~Zheng, X.~Tang, X.~Wei, H.~Shen, and L.~Zhao, ``Channel assignment for
  hybrid noma systems with deep reinforcement learning,'' \emph{IEEE Wireless
  Communications Letters}, vol.~10, no.~7, pp. 1370--1374, 2021.

\bibitem{liu2024dnn}
Z.~Liu, H.~Du, J.~Lin, Z.~Gao, L.~Huang, S.~Hosseinalipour, and D.~Niyato,
  ``{DNN Partitioning, Task Offloading, and Resource Allocation in Dynamic
  Vehicular Networks: A Lyapunov-Guided Diffusion-Based Reinforcement Learning
  Approach},'' \emph{arXiv preprint arXiv:2406.06986}, 2024.

\bibitem{9964376}
C.~Chaieb, F.~Abdelkefi, and W.~Ajib, ``Deep reinforcement learning for
  resource allocation in multi-band and hybrid oma-noma wireless networks,''
  \emph{IEEE Transactions on Communications}, vol.~71, no.~1, pp. 187--198,
  2023.

\bibitem{de2018molgan}
N.~De~Cao and T.~Kipf, ``{MolGAN: An Implicit Generative Model for Small
  Molecular Graphs},'' \emph{arXiv preprint arXiv:1805.11973}, 2018.

\bibitem{simonovsky2018graphvae}
M.~Simonovsky and N.~Komodakis, ``{Graphvae: Towards Generation of Small Graphs
  Using Variational Autoencoders},'' in \emph{Artificial Neural Networks and
  Machine Learning--ICANN 2018: 27th International Conference on Artificial
  Neural Networks, Rhodes, Greece, October 4-7, 2018, Proceedings, Part I
  27}.\hskip 1em plus 0.5em minus 0.4em\relax Springer, 2018, pp. 412--422.

\bibitem{yun2019graph}
S.~Yun, M.~Jeong, R.~Kim, J.~Kang, and H.~J. Kim, ``{Graph Transformer
  Networks},'' \emph{Advances in neural information processing systems},
  vol.~32, 2019.

\bibitem{vignac2022digress}
C.~Vignac, I.~Krawczuk, A.~Siraudin, B.~Wang, V.~Cevher, and P.~Frossard,
  ``Digress: Discrete denoising diffusion for graph generation,'' \emph{arXiv
  preprint arXiv:2209.14734}, 2022.

\bibitem{ray2020predicting}
S.~Ray, S.~Lall, A.~Mukhopadhyay, S.~Bandyopadhyay, and A.~Sch{\"o}nhuth,
  ``{Predicting Potential Drug Targets and Repurposable Drugs for Covid-19 via
  A Deep Generative Model for Graphs},'' \emph{arXiv preprint
  arXiv:2007.02338}, 2020.

\bibitem{9625800}
Y.~Yuan, Z.~Li, Z.~Liu, Y.~Yang, and X.~Guan, ``{Double Deep Q-Network Based
  Distributed Resource Matching Algorithm for D2D Communication},'' \emph{IEEE
  Transactions on Vehicular Technology}, vol.~71, no.~1, pp. 984--993, 2022.

\end{thebibliography}


\end{document}